\documentclass[conference]{IEEEtran}
\IEEEoverridecommandlockouts
\usepackage{cite}
\usepackage{amsmath,amssymb,amsfonts}
\usepackage{algorithmic}
\usepackage{graphicx}
\usepackage{textcomp}
\makeatletter

\newcommand{\Rmnum}[1]{\expandafter\@slowromancap\romannumeral #1@}
\makeatother
\usepackage{multirow, makecell}
\def\BigRoman{\uppercase\expandafter{\romannumeral\number\count 255 }}
\def\Romannumeral{\afterassignment\BigRoman\count255=}
\usepackage{xcolor}
\usepackage{multirow}
\usepackage{array}
\usepackage{tikz}
\def\BibTeX{{\rm B\kern-.05em{\sc i\kern-.025em b}\kern-.08em
    T\kern-.1667em\lower.7ex\hbox{E}\kern-.125emX}}
\begin{document}

\title{Design of an EEG-based Drone Swarm Control System using Endogenous BCI Paradigms
\footnote{{\thanks{20xx IEEE. Personal use of this material is permitted. Permission
from IEEE must be obtained for all other uses, in any current or future media, including reprinting/republishing this material for advertising or promotional purposes, creating new collective works, for resale or redistribution to servers or lists, or reuse of any copyrighted component of this work in other works.

This work was supported by Defense Acquisition Program Administration (DAPA) and Agency for Defense Development (ADD) of Korea.}
}}
}

\author{\IEEEauthorblockN{Dae-Hyeok Lee}
\IEEEauthorblockA{\textit{Dept. Brain and Cognitive Engineering}\\
\textit{Korea University} \\
Seoul, Republic of Korea \\
lee\_dh@korea.ac.kr}\\

\IEEEauthorblockN{Hyung-Ju Ahn}
\IEEEauthorblockA{\textit{Dept. Brain and Cognitive Engineering}\\
\textit{Korea University} \\
Seoul, Republic of Korea \\
hj\_ahn@korea.ac.kr}\\

\and

\IEEEauthorblockN{Ji-Hoon Jeong}
\IEEEauthorblockA{\textit{Dept. Brain and Cognitive Engineering}\\
\textit{Korea University} \\
Seoul, Republic of Korea \\
jh\_jeong@korea.ac.kr}\\

\IEEEauthorblockN{Seong-Whan Lee}
\IEEEauthorblockA{\textit{Dept. Artificial Intelligence}\\
\textit{Korea University} \\
Seoul, Republic of Korea \\
sw.lee@korea.ac.kr}
}


\maketitle

\begin{abstract}
Non-invasive brain-computer interface (BCI) has been developed for understanding users' intentions by using electroencephalogram (EEG) signals. With the recent development of artificial intelligence, there have been many developments in the drone control system. BCI characteristic that can reflect the users' intentions led to the BCI-based drone control system. When using drone swarm, we can have more advantages, such as mission diversity, than using a single drone. In particular, BCI-based drone swarm control could provide many advantages to various industries such as military service or industry disaster. BCI Paradigms consist of the exogenous and endogenous paradigms. The endogenous paradigms can operate with the users' intentions independently of any stimulus. In this study, we designed endogenous paradigms (i.e., motor imagery (MI), visual imagery (VI), and speech imagery (SI)) specialized in drone swarm control, and EEG-based various task classifications related to drone swarm control were conducted. Five subjects participated in the experiment and the performance was evaluated using the basic machine learning algorithm. The grand-averaged accuracies were 51.1\% ($\pm$ 8.02), 53.2\% ($\pm$ 3.11), and 41.9\% ($\pm$ 6.09) in MI, VI, and SI, respectively. Hence, we confirmed the feasibility of increasing the degree of freedom for drone swarm control using various endogenous paradigms.
\end{abstract}

\begin{small}
\textbf{\textit{Keywords- brain-computer interface; electroencephalogram; drone swarm control; intuitive paradigm}}\\
\end{small}

\section{Introduction}
Brain-computer interface (BCI) uses brain waves to identify users' intentions and control robots or computers accordingly. Non-invasive BCI technology \cite{lee2015subject, nguyen2019adaptive, won2017motion, choi2020application, lee2020continuous} allows users to communicate with external devices without brain implant surgery. Non-invasive BCI systems have been applied for interaction using a robotic arm \cite{penaloza2018bmi, jeong2020multimodal}, a wheelchair \cite{li2019eeg, kim2018commanding}, and a speller \cite{gu2019online, lee2018high}.

Recently, non-invasive BCI systems for commercial and military applications are being actively researched. One of the most challenging research topics is collaborating with a swarm of robots or drones using electroencephalogram (EEG) signals \cite{karavas2017hybrid, jeong2020towards, choi2020application}. Applications of drone swarm have great potentials for both military and civilian usages. Rojas \textit{et al}.\cite{rojas2020electroencephalographic} proposed EEG indicators for objective assessment of cognitive workload during the teleoperation of an unmanned aerial vehicle shepherding a swarm of unmanned ground vehicles. Karavas \textit{et al}.\cite{karavas2017hybrid} demonstrated a hybrid control of drone swarm using joystick and EEG.

Over the past decades, BCI has been developed with various kinds of paradigms. BCI paradigms can be approximately sorted into exogenous and endogenous paradigms. The exogenous BCI paradigms require external stimuli such as LED flicker \cite{kwak2017convolutional, lee2018high}. Motor imagery (MI), visual imagery (VI), and speech imagery (SI) are well-known endogenous BCI paradigms, which can identify users' intentions without actual execution. Many studies of the EEG-based drone control systems have adopted the exogenous BCI paradigm since it has advantages for minimal training and easy experimental setup \cite{kwak2017convolutional}. The endogenous paradigms have the advantage that users' intentions are well reflected independently of stimuli. Endogenous paradigms allow users to imagine the tasks intuitively, which occur less fatigue when performing the tasks \cite{jeong2020multimodal, kosmyna2018attending, nguyen2017inferring}. Nguyen \textit{et al}. \cite{nguyen2019adaptive} proposed the system to control the swarm density and the shape of the formation with a 4-degree of freedom using MI and SI paradigms. Koizumi \textit{et al}. \cite{koizumi2018development} investigated to control a single quadcopter's 3-D movement with a 6-degree of freedom based on VI and SI paradigms. However, many related studies have a critical limitation about the low degree of freedom which hinders the practicality of the BCI system.

\begin{figure}[t]
\centerline{\includegraphics[width=\columnwidth]{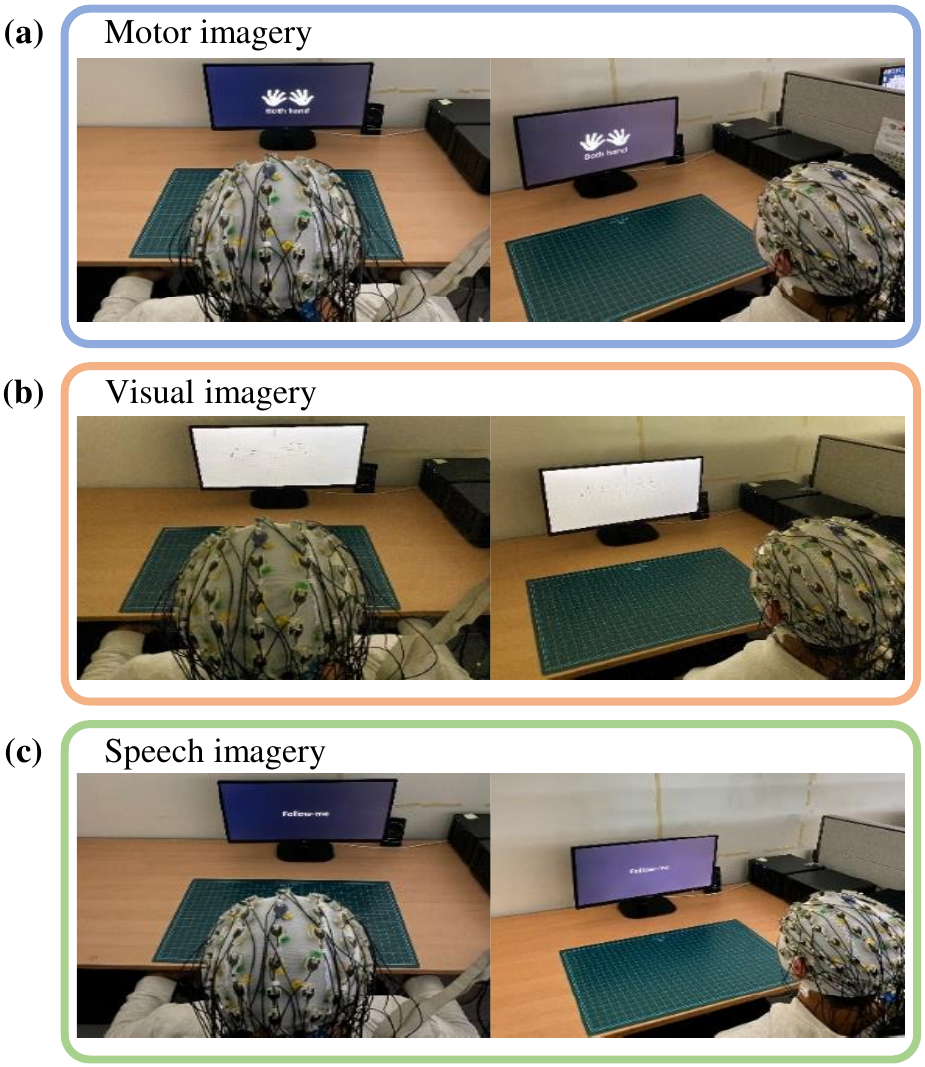}}
\caption{Experimental environments for EEG data acquisition in each paradigm. Subjects receive various tasks corresponding to each paradigm through a monitor. (a) MI, (b) VI, (c) SI.}
\end{figure}

In this study, we measured EEG signals using our designed endogenous paradigms (i.e., MI, VI, and SI). The classes used in the MI paradigm consist of the most essential direction control (i.e., left, right, up, and down). We designed the VI paradigm for the most basic formation control (i.e., fall-in, spread-out, and split). The SI paradigm we designed was for useful high-level mission control (i.e., go, stop, follow me, and return). To the best of our knowledge, this is the first study to utilize various endogenous paradigms for increasing the degree of freedom for drone swarm control and to demonstrate the feasibility of classifying the commands assigned to each paradigm. Also, we achieved robust classification performance in the commands assigned to each paradigm compared with the chance-level accuracies (i.e., MI: 0.25, VI: 0.33, and SI: 0.25).

The rest of this paper is organized as follows. Section {\Romannumeral 2} gives a description of the experimental protocols, EEG signals acquisition, system architecture, and data analysis. Section {\Romannumeral 3} presents the results of performance accuracies for classifying the commands assigned to each paradigm and discussion about our study. In session {\Romannumeral 4}, the conclusion and future works are described.\\

\section {Materials and Methods}
\subsection{Subjects}
Five healthy subjects (S1-5, 3 males and 2 females, aged 25.4 ($\pm 3.1$)) volunteered in the experiments. All experiments were approved by the Institutional Review Board at Korea University (KUIRB-2020-0318-01). Before the experiment, we informed them to get adequate sleep (over 7 hr.) and avoid any alcohol the day before the experiment. Subjects were informed about the experimental protocols and procedures. They provided their written consent according to the Declaration of Helsinki.

\subsection{Experimental Paradigms}
We designed three endogenous paradigms for increasing the degree of freedom for drone swarm control based on the conventional studies \cite{jeong2020multimodal, kosmyna2018attending, nguyen2017inferring}.
\begin{figure}[t!]
\centering
\scriptsize
\centerline{\includegraphics[width=\columnwidth]{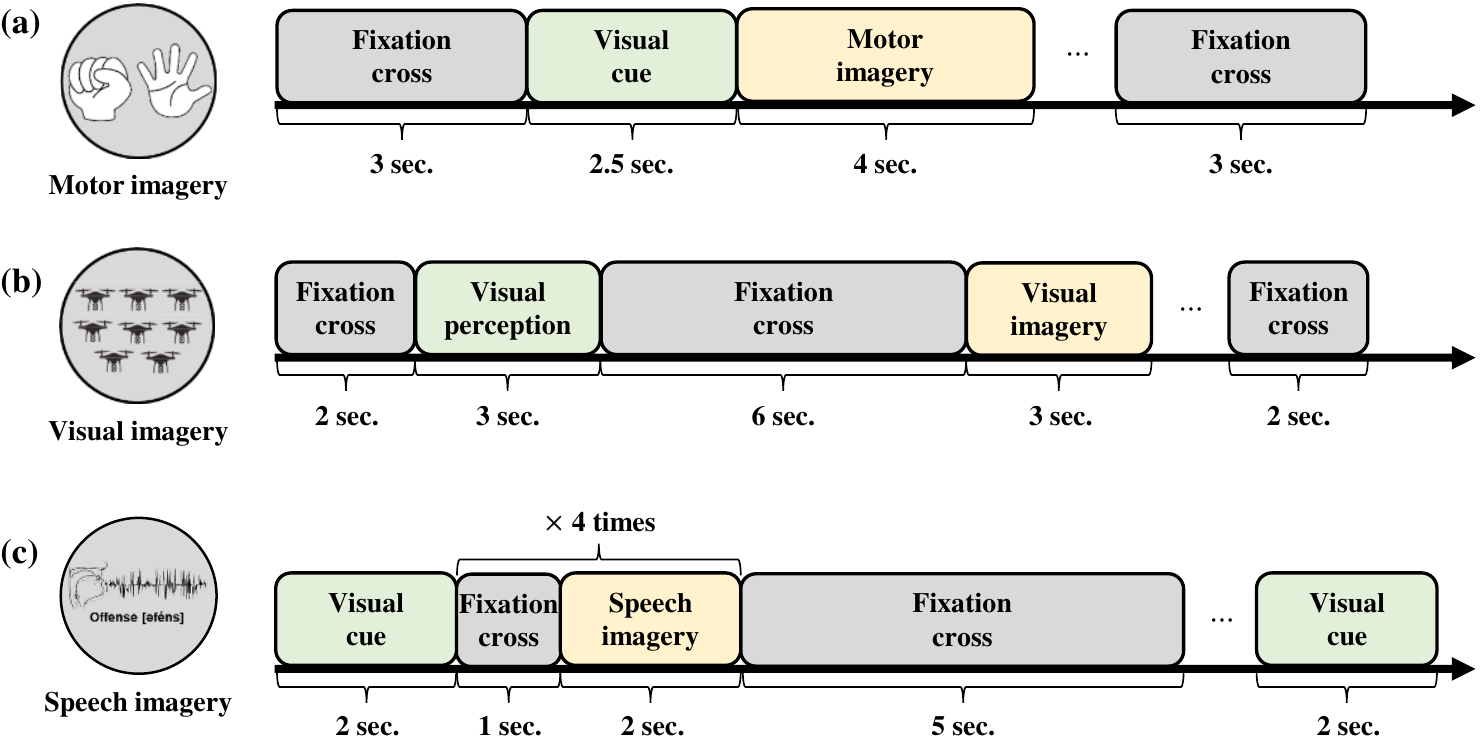}}
\caption{Experimental paradigms for acquiring EEG data-related various drone swarm tasks control. The fixation cross was used for eliminating any possible afterimages. (a) MI, (b) VI, (c) SI.}
\end{figure}
\begin{figure*}[t]
\centerline{\includegraphics[width=\textwidth]{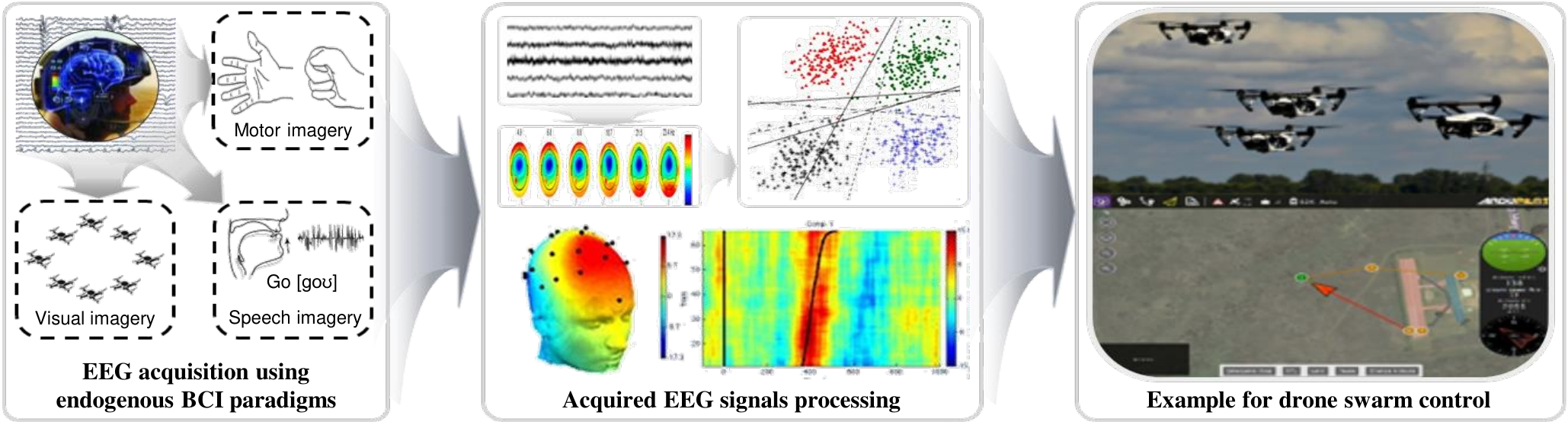}}
\caption{System architecture for increasing the degree of freedom for drone swarm control. The degree of freedom of drone swarm control is increased by analyzing the acquired EEG data through various endogenous BCI paradigms using machine learning methods.}
\end{figure*}

\subsubsection{Motor imagery}
MI tasks include left hand, right hand, both hands, and foot. Left and right hand correspond to commands for the left and right direction of drone swarm. Both hand and foot correspond to the high and low altitude flight missions, respectively. The experimental protocol was composed of three types of visual cues. The fixation cross was presented for 3 sec. at the start of each trial. Instruction for each MI task was given for 2.5 sec. with an image of feet or hand (left, right, or both) randomly. A blank screen was presented for 4 sec. to perform MI tasks. 50 trials were acquired for each class (a total of 200 trials). See Fig. 1(a) and Fig. 2(a) for more details.

\subsubsection{Visual imagery}
VI paradigm was designed to control the cohesion levels and formations of a swarm of drones. Three different tasks (i.e., increasing swarm cohesion level (fall-in), decreasing swarm cohesion level (spread-out), and separated into two groups (split)) were provided to the subjects in video format. As shown in Fig. 1(b) and Fig. 2(b), the presentation sequence is comprised of four stages. The fixation cross was presented for 2 sec.. A video of different swarm behavior for each task was showed randomly, and it lasted for 3 sec.. The fixation cross was presented for 6 sec. to eliminate any possible afterimages. A blank image is presented for 3 sec. for the visual imagery task. 50 trials were acquired for each class (a total of 150 trials).

\subsubsection{Speech imagery}
Four words (i.e., Go, Stop, Follow-me, and Return) were selected for the SI tasks. ‘Go’ is a command to start or restart the mission. ‘Stop’ is an instruction to pause the mission. ‘Follow-me’ is to help the user by moving to the user's current point, no matter what the interference may be. ‘Return’ can be used to return drone swarm to base. The experiment process consisted of three phases. The visual cue of each word was presented for 2 sec. randomly. The visual cue was provided over 2 sec. regardless of the word length. A fixation cross was displayed for 1 sec. and a blank image was displayed for 2 sec.. Subjects were instructed to perform a SI task when a blank image was presented. The subject repeated the SI task and fixation cross trial 4 times per visual cue. The last of four fixation cross trials was lasted for 5 sec. to eliminate the afterimages. 50 trials were acquired for each class (a total of 200 trials). See Fig. 1(c) and Fig. 2(c) for more details.

\subsection{EEG Signal Acquisition}
We measured the EEG signals in terms of various kinds of drone swarm control scenarios using BrainVision Recorder (BrainProducts GmbH, Germany). EEG signals which were measured in 3 endogenous paradigms were acquired using 64 Ag/AgCl electrodes, and all electrodes were placed on the subjects’ scalp according to the international 10/20 system. The reference electrode was placed at FCz, and the ground electrode was placed at FPz. We set up the sampling rate to 1,000 Hz and a 60 Hz notch filter was applied. Before the acquisition of EEG data, the impedance of all electrodes was kept below 10 k$\Omega$ by injecting conductive gel.

\subsection{System Architecture}
As shown in Fig. 3, the EEG-based drone swarm control system consists of three parts: EEG acquisition, signal processing, and drone swarm control. We adopted the ROS (Robotic Operating System) to communicate MATLAB with the drone swarm control module. The EEG data were acquired using three endogenous BCI paradigms. Significant features were extracted from acquired EEG signals using MATLAB and the classification of various classes was carried out through the basic machine learning algorithm. We designed the endogenous paradigms for a highly-utilized BCI-based drone swarm control system.

\subsection{Data Analysis}
The acquired EEG signals were down-sampled from 1,000 to 100 Hz. They were preprocessed by using a bandpass filter with a zero-phase, 2nd Butterworth filter. In the imagination decoding from EEG signals, mu and beta bands were usually used. We applied an independent component analysis which is one of the most used preprocessing techniques to remove the artifacts of EEG signals such as eye blinks for obtaining clean EEG data. We segmented the data into 4 s epochs for each trial. We applied a common spatial pattern (CSP) \cite{ang2008filter} for extracting informative spatial features. We used a linear discriminant analysis (LDA) \cite{cho2018classification} for classifying various classes using the one-versus-rest strategy. EEG signal processing was conducted using a BBCI toolbox \cite{blankertz2010berlin} in MATLAB 2019a environment. We applied 5-fold cross-validation to evaluate the classification performance fairly. Also, we repeated the 5-fold cross-validation five times.

\section {Results and Discussion}
The classification accuracies for each paradigm were calculated via the CSP-LDA algorithm. The grand-averaged accuracies are the average values of accuracy with five times repeated 5-fold cross-validation. Fig. 4 shows the accuracies of all subjects for each paradigm.

In the case of MI, subject 3 (S3) represented the highest classification accuracy as 63.8\%, but subject 4 (S4) showed the lowest accuracy as 39.1\%. However, all subjects indicated higher accuracies than chance level accuracy for classifying 4-class (approximately 25.0\%). The grand-averaged accuracy of the MI experiment was 51.1\% ($\pm$ 8.02). 

In the VI experiment, subject 4 (S4) showed the highest performance as 59.8\%, and subject 1 (S1) showed a performance that was almost the same as subject 4 (S4). On the other hand, subject 3 (S3) indicated the lowest performance as 42.1\%. As with the MI experimental results, all subjects represented higher performances than chance level accuracy for classifying 3-class (approximately 33.3\%). The grand-averaged performance of the VI experiment was 53.2\% ($\pm$ 3.11). 

\begin{figure}[t]
\centerline{\includegraphics[scale=0.8]{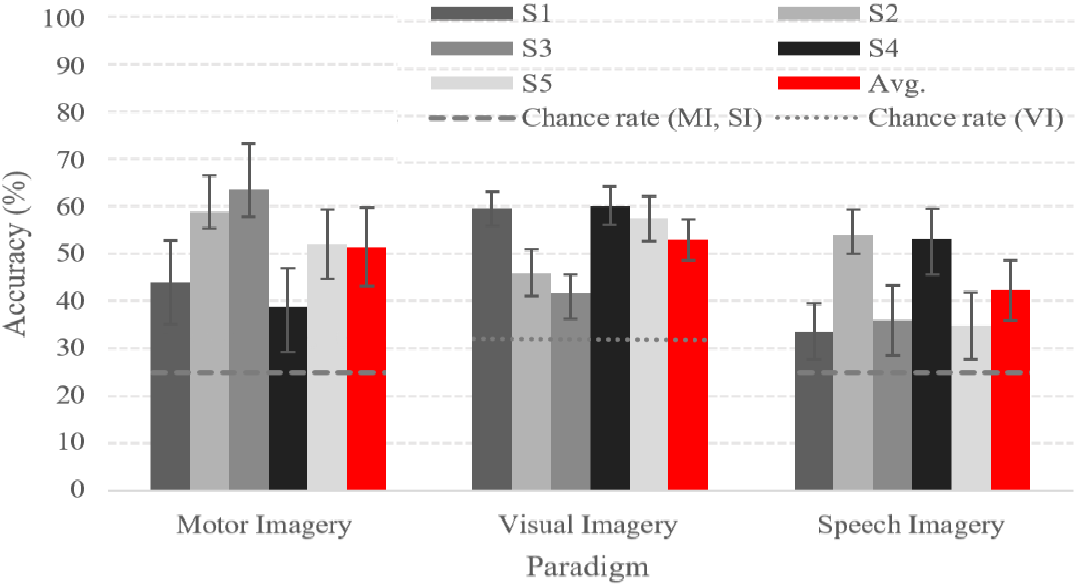}}
\caption{Classification accuracies of each endogenous paradigm across all subjects. The red-colored bar graphs represent the grand-averaged accuracy of all subjects in each paradigm.}
\end{figure}

Finally, in the case of SI, subject 2 (S2) indicated the highest accuracy as 53.9\%, and subject 4 (S4) indicated an accuracy that was almost similar to subject 2 (S2). Meanwhile, subject 1 (S1) represented the lowest accuracy as 33.6\%, and subject 3 (S3) and subject 5 (S5) represented similar accuracies as subject 1 (S1). In the SI experiment, all subjects showed higher accuracies than chance level accuracy for classifying 4-class (approximately 25.0\%). The grand-averaged accuracy of the SI experiment was 41.9\% ($\pm$ 6.09). 

For the purpose of evaluating classification performance, although we applied the basic machine learning algorithm, all subjects showed higher accuracies than the chance level accuracy in all paradigms. Through this, we could confirm that the EEG data acquired in each paradigm were high-quality data. Also, in each paradigm, several subjects showed poor performances, and we think it is a phenomenon caused by the difficulty of performing the image task of the paradigms. Through experiments of various paradigms, we could confirm that intuitive instructions similar to the real-world environments, such as control of the actual drone swarm, are needed. It could be more helpful for subjects to carry out intuitive instructions.

\section{Conclusion and Future works}
In this paper, we designed the endogenous paradigms (i.e., MI, VI, and SI), and by utilizing these paradigms, we acquired high-quality EEG data. All subjects successfully carried out drone swarm control-related tasks through each paradigm, and the corresponding tasks can contribute to increasing the degree of freedom for drone swarm control. Since our study is to check the feasibility of increasing the degree of freedom for drone swarm control using various endogenous paradigms, the basic machine learning algorithm was used for EEG classification and its performances are slightly higher than the chance level accuracies.

Hence, we will design a deep learning network model that uses the EEG signal acquired at endogenous paradigms as input to make higher performances. In order to apply to real-world environments robustly, we will develop the systems that can set modes to their respective paradigms, freeing mode conversion.

\section{Acknowledgement}
The authors thank to B.-H. Kwon for his help with the design of visual imagery paradigm.\\
\bibliographystyle{IEEEbib}
\bibliography{refs}

\end{document}